\documentclass[]{agujournal}
\draftfalse

\journalname{JGR-Oceans}

\usepackage{graphicx}
\usepackage{color}
\usepackage{multirow}
\usepackage{amsmath,amsfonts,amssymb,mathtools}   

\newcommand{\mathsout}[1]
{\bgroup\mathchoice
  {\sbox0{$\displaystyle{#1}$}%
    \usebox0\hspace{-\wd0}%
    \rule[0.5\ht0-0.5\dp0-.5pt]{\wd0}{1pt}}%
  {\sbox0{$\textstyle{#1}$}%
    \usebox0\hspace{-\wd0}%
    \rule[0.5\ht0-0.5\dp0-.5pt]{\wd0}{1pt}}%
  {\sbox0{$\scriptstyle{#1}$}%
    \usebox0\hspace{-\wd0}%
    \rule[0.5\ht0-0.5\dp0-.5pt]{\wd0}{1pt}}%
  {\sbox0{$\scriptscriptstyle{#1}$}%
    \usebox0\hspace{-\wd0}%
    \rule[0.5\ht0-0.5\dp0-.5pt]{\wd0}{1pt}}%
\egroup}

\begin{document}

\title{
A framework for modelling linear surface waves on shear currents 
in slowly varying waters
}

\authors{Yan Li\affil{1,2} and Simen {\AA}. Ellingsen\affil{1} }

\affiliation{1}{Department of Energy and Process Engineering, Norwegian University of Science and Technology, Trondheim, Norway}
\affiliation{2}{Department of Engineering Science, University of Oxford, Oxford, United Kingdom}

\correspondingauthor{Y. Li} {yan.li@ntnu.no}


\begin{keypoints}
\item A framework for analysing waves atop currents of arbitrary vertical shear, allowing  slow horizontal depth and current variation;
\item Direct integration method for simple and fast evaluation of dispersion and wave action;
\item Favourable to existing numerical and analytical approaches for a wide range of practical applications.
\end{keypoints}

\begin{abstract}
We present a theoretical and numerical framework -- which we dub the Direct Integration Method (DIM) -- for simple, efficient and accurate evaluation of surface wave models allowing presence of a current of arbitrary depth dependence, and where bathymetry and ambient currents may vary slowly in horizontal directions.
On horizontally constant water depth and shear current the DIM numerically evaluates the dispersion relation of linear surface waves to arbitrary accuracy, and we argue that for this purpose it is superior to two existing numerical procedures: the piecewise-linear approximation and a method due to \textit{Dong \& Kirby} [2012].
The DIM moreover yields the full linearized flow field at little extra cost. We implement the DIM numerically with iterations of standard numerical methods. The wide applicability of the DIM in an oceanographic setting in four aspects is shown.
Firstly, we show how the DIM allows practical implementation of the wave action conservation equation recently derived by \textit{Quinn et al.} [2017]. Secondly, we demonstrate how the DIM handles with ease cases where existing methods struggle, i.e.\ velocity profiles $\mathbf{U}(z)$ changing direction with vertical coordinate $z$, and strongly sheared profiles. Thirdly, we use the DIM to calculate and analyse the full linear flow field beneath a 2D ring wave upon a near--surface wind--driven exponential shear current, revealing striking qualitative differences compared to no shear. Finally we demonstrate that the DIM can be a real competitor to analytical dispersion relation approximations such as that of \textit{Kirby \& Chen} [1989] even for wave/ocean modelling. 
\end{abstract}

\section{Introduction}

Surface waves in ocean and coastal waters are often affected by currents. Particularly when the current varies with depth --- i.e., it has vertical shear --- the interactions between surface waves and current can be rich and highly non--trivial even in the linear wave regime. While this has been recognized for a long time \citep[e.g.][]{peregrine76}, it is recently becoming increasingly clear that the effect of shear in the water column must be accounted for in order that environmental waves may be fully understood, and adequately modeled, 
as emphasized in a recent review of coastal wave modelling \citep{cavaleri18}. 
Several oceanographic models such as Delft-3D \citep{elias12} and ROMS, used for example in the coupled COAWST model \citep{kumar12}, now include as an option of the wave--dispersion correction due to a horizontal ambient current $\mathbf{U}(z)$ which varies with vertical coordinate $z$. Necessary theoretical tools and insights to this end have been developed in recent studies \citep{banihashemi17, quinn17}, where it was concluded that wrongly accounting for shear in such wave models, or neglecting it, can lead to serious errors. Failure to include Langmuir turbulence, a direct consequence of wave--shear current interaction \citep{leibovich83}, is a prime suspect for the systematic misprediction in global climate models of the ocean surface 
temperature and boundary layer depth, particularly in the southern oceans \citep{belcher12}. 
Wave effects moreover greatly influence storm surge inundation as shown by \citet{wu18}, where strongly sheared currents are present.

Knowledge of the effect of shear on wave dispersion is also necessary in order to remotely measure near--surface currents using 
  X-band of HF/VHF radar   \citep{stewart74,teague86,shrira01,lund15,campana17}, 
used \emph{inter alia} to shed light on exchange of heat, mass and momentum between ocean and atmosphere, and transportation of nutrients and pollutants.

In this article we present a new framework for numerical calculation of wave dispersion on arbitrary shear currents, which 
compliments the analytical framework developed by \citet{ellingsen17}, and applied to the case of ship waves by \citet{li17a}. We refer to it as the Direct Integration Method (DIM). The DIM is simple to implement, combining only standard operations for solving linear inhomogeneous differential equations, numerical integration and root--finding, for which any of a number of methods may be used. The implementation tested herein with an iterative scheme which 
uses very basic procedures: finite differences, Simpson's method, and Newton's method, respectively (an example implementation in MATLAB is included in supplementary materials). 
In the interest of fair comparison the iterative numerical scheme is deliberately simple, and 
we do not claim it is the universally optimal option for implementation of the DIM.

The DIM has a number of attractive features. It can handle ambient currents $\mathbf{U}(z)$ ($z$: vertical coordinate) 
which change direction with depth with the same ease as unidirectional currents. It facilitates estimation 
of the relative error of the calculated value of $c(\mathbf{k})$ at little additional cost ($\mathbf{k}$: wave vector in the horizontal plane; $c$: phase velocity). And it may provide the full sub--surface flow field within linear wave theory, without any increase in complexity. 
The general solutions of the full flow field are the fundamental components of analysing 3D rotational waves of finite amplitude, a question for future studies. 

A particular feature of the DIM is its ability to include the effects of slowly varying water depth. 
A wave action conservation equation for this situation was recently derived by \citet{quinn17}, but those authors deemed that its application in ocean and climate models was impractical due to computational cost. We believe that with the DIM this could change radically. We present both an analytical framework and an extension of DIM numerics to evaluate the equation of \citet{quinn17}. 

We argue that the DIM is superior to the two existing 
numerical 
methods for calculating $c(\mathbf{k})$ with arbitrary accuracy over constant water depth that we are aware of, namely the piecewise--linear method \citep[e.g.][]{zhang05}, and a method due to \citet{dong12}. It is considerably simpler to implement than the former and is faster and easier to parallelize than the latter, which also cannot provide the flow field in a straightforward manner. Neither method has been developed to handle changing depth.

The structure of the paper is as follows. In Sec.~\ref{sec:def} we define the system under consideration. Sec~\ref{sec:appr} reviews existing methods for calculating or estimating $c(\mathbf{k})$ on a current $\mathbf{U}(\mathbf{k})$ of arbitrary depth dependence, including numerical procedures of arbitrary accuracy, and widely used analytical approximations. The Direct Integration Method is presented in Sec.~\ref{sec:DIM}. 
We apply it with an iterative scheme to a range of different cases in Sec.~\ref{sec:num}. 
In Sec.~\ref{sec:comparison} we compare the DIM 
to existing numerical and analytical approaches in terms of accuracy and cost, before concluding remarks are provided in Sec.~\ref{sec:concl}. The 
numerical performance
of the iterative scheme
is tested in an appendix.


\section{Theory formalism}\label{sec:def}

\begin{figure}
	\graphicspath{{figures/}}
	\centering{\includegraphics[width=.75\textwidth]{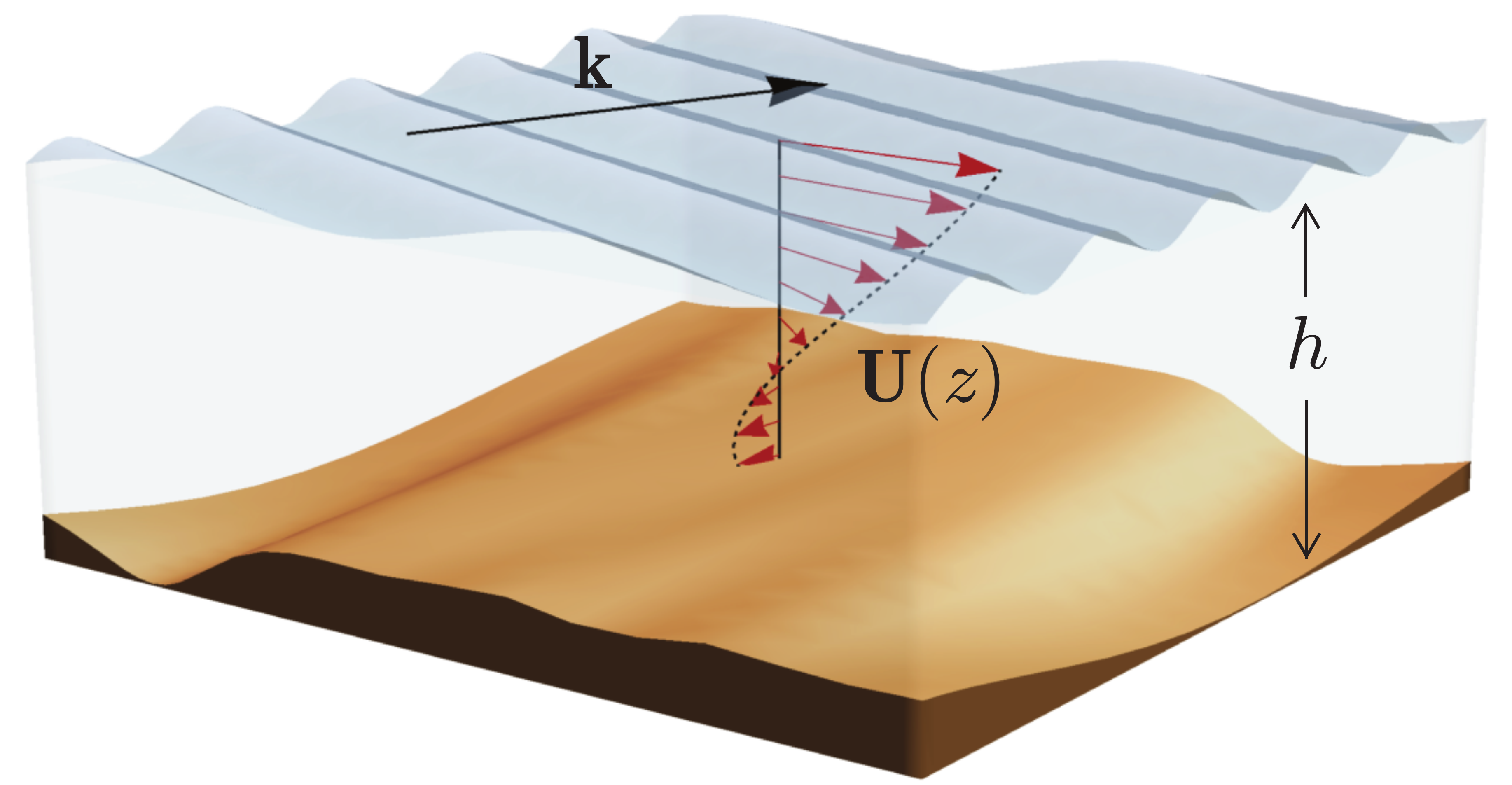}}
	\label{fig:geometry}
	\caption{Geometry of the three-dimensional wave and current system. }
\end{figure}

We consider a plane wave running on a horizontal background flow $\mathbf{U}(\mathbf{x},z)$ where $ \mathbf{x}=(x,y) $ is the position vector in the horizontal plane and $z$ is the vertical axis such that  the undisturbed surface is located at $z=0$, over varying water depth $ h(\mathbf{x})$.

A geometry of the system is depicted in Fig.~\ref{fig:geometry}. Assume incompressible and inviscid flow, and that the medium above the surface can be neglected. We assume perturbations of the background flow due to the wave motion is small, and work eventually within linear wave theory. A wave in the horizontal plane with wave vector $\mathbf{k} = (k_x, k_y) = k(\cos\theta,\sin\theta)$ generates perturbations that are understood to be proportional to $ \exp(\mathrm{i} \mathbf{k}\cdot\mathbf{x}-\mathrm{i} kc(\mathbf{k}) t) $, where 
$k=|\mathbf{k}|$ is the wave number, $c(\mathbf{k})$ is the phase velocity of the wave along direction $\mathbf{k}$, and $t$ is the time.

Before proceeding to the governing equations and boundary conditions, we first introduce two key assumptions;
\begin{enumerate}
	\item[I] Fast variation of wave phase; the wave phase $S\equiv \mathbf{k}\cdot\mathbf{x}-kct $ is of rapid spatial and time variation -- $ \mathcal{O}(1) $ -- when compared to the slow variation of the shear current, water depth 	, and wave amplitude $ A$ in the horizontal plane and time. 
	\item[II] We assume 
	wave motions to be affected by a background current, but not vice versa. 
\end{enumerate}

Assumption II allows us to first look at purely non-wave motions (wherein parameters are defined with the superscript `$ ^{(0)}$') and then together with wave motions. 

\subsection{The steady background flow}
When there is no wave motion and 
the time dependence
is negligible 
, the Euler and continuity equations yield 
after eliminating the horizontal velocity perturbation components $\hat{u}^{(0)}$ and $\hat{v}^{(0)}$,

\begin{subequations} \label{eq:gvn-nonW}
\begin{align}
\nabla \cdot \mathbf{U} + {\hat{w}}'^{(0)} = & 0; \\
(\mathbf{U}\cdot \nabla) \mathbf{U} +  \mathbf{U}' {\hat{w}}^{(0)} = & -\nabla {\hat{p}}^{(0)}/\rho; \\
(\mathbf{U}\cdot \nabla) {\hat{w}}^{(0)} + {\hat{w}}'^{(0)} =& - {\hat{p}}'^{(0)}/\rho. 
\end{align}
\end{subequations}
where a
prime denotes the derivative with respect to $ z $, $ \hat{w}^{(0)} $ is the vertical velocity and $ \hat{p}^{(0)} $ is the dynamic pressure, 
 $g$ is the grativational acceleration, 
and the operator $ \nabla \equiv (\partial_x, \partial_y) $. 
Eqs.~\eqref{eq:gvn-nonW} assume no variation of density $\rho$, precluding stratification and internal waves. 

Thelinearized boundary conditions are 
\begin{subequations} \label{eq:bc-nonW}
\begin{align}
    \hat{p}^{(0)}-\rho g \eta^{(0)} = & 0; ~\text{at}~ z=0, \\
   \hat{w}^{(0)} - \mathbf{U}\cdot \nabla\eta^{(0)} = & 0; ~\text{at}~ z=0, \\
   \hat{w}^{(0)}  - \mathbf{U}\cdot \nabla h =& 0; \text{at}~ z=-h,
\end{align}
\end{subequations}
where $ \eta^{(0)} $ is the surface elevation
and $\rho$ is the density of fluid. 

The solutions of \eqref{eq:gvn-nonW} and \eqref{eq:bc-nonW} work as boundary conditions when additional wave motions are considered. 

\subsection{Small wave motion in the presence of a shear current}
We now proceed to consider wave motion together with a background current. A  generic field variable $ \tilde{\chi}$  is then expressed 
$ \tilde{\chi}\equiv \hat{\chi}^{(0)}+ {\chi} \exp[\mathrm{i}\mathbf{k}\cdot\mathbf{x}-\mathrm{i} kc(\mathbf{k})t]$, 
the latter of which is the variable due to wave motion. 

The flow of a wave-shear current system is governed by the Euler and continuity equations. Assumption I allows necessary linearisation as follows.  After the linearisation with respect to 
surface steepness 
$ \mathrm{\epsilon} = kA $
($ \mathrm{\epsilon}\ll 1 $
, $A$ is a characteristic wave amplitude) 
the flow may be expressed as the boundary value problem 
\begin{subequations} \label{eq:govn}
\begin{align}
(\mathbf{k} \cdot \mathbf{U}-kc)(w''-k^2w) = { \mathbf{k}\cdot\mathbf{U}''}w;~~ &-h<z<\eta^{(0)} , \label{rayleigh}
\\
(\mathbf{k} \cdot \mathbf{U}-kc)^2 w'- (gk^2+\frac{T k^4}{\rho})w- \mathbf{k} \cdot \mathbf{U}'(\mathbf{k} \cdot \mathbf{U}-kc) w = 0;~~ &z=\eta^{(0)} , \label{BC}\\
w =  0; ~~&z=-h(x,y),
\end{align}
\end{subequations}
where $w(\mathbf{k},z)$ is the amplitude of the vertical velocity due to wave motion
and $T$ is the surface tension coefficient. 
Eq.~\eqref{rayleigh} is called the Rayleigh equation (or the inviscid Orr--Sommerfeld equation). We are thus faced with an eigenvalue problem with two unknowns, $w(z)$ and $c(\mathbf{k})$. 
%

\section{Existing approaches for constant $h$} \label{sec:appr}
Existing approaches for calculating or estimating the value of $c(\mathbf{k})$ 
when $h$ is constant with respect to $\bf{x}$ 
may be divided into two categories: numerical procedures with arbitrary accuracy, and analytical approximations with theoretical error. In the following we briefly review these 
wherein constant water depth is assumed and $ \eta^{(0)}\equiv 0 $ is hence obtained.

\subsection{Arbitrary accuracy}\label{sec:arbacc}

Numerical schemes to determine $c(\mathbf{k})$ in the past have included, in particular, two very different strategies. 

The first, and oldest approach is to divide the water column into $n$ artificial layers, and presume $\mathbf{U}(z)$ to vary as a linear function of $z$ within each layer. The linearised Euler equations now permit explicit solutions with undetermined coefficients within each layer. When $U(z)$ does not vary in direction there are $n+1$ free coefficients which are determined as zeroes of the system determinant. We refer to this method as the 
piecewise--linear approximation (PLA). 
The idea goes back well 
over a century \citep{rayleigh1892}, and has recently been analysed in further detail \citep{zhang05,smeltzer17}. 
The method is tried and trusted, physically intuitive, and reasonably efficient when moderate accuracy is required (4-5 layers are typically sufficient for error $<5\%$), but has certain drawbacks. 
The foremost of these is that $n+1$ eigenvalues for $c(\mathbf{k})$ are found of which two must be chosen which describe surface wave propagation, the rest being vorticity waves generated by the artificial discontinuities of $U'(z)$ and should be discarded. This considerably complicates implementation. Secondly, directly generalizing $U(z)$ to allow changing direction would double the number of undetermined coefficients, much increasing the cost. A more sophisticated procedure could likely avoid this, but we are not aware of any implementation of the PLA for turning $\mathbf{U}(z)$ to date.

An alternative procedure was used by \cite{dong12}. They introduce an additional function $ Q(z) = w(z)/w'(z)$ which transforms the Rayleigh equation into a nonlinear ordinary differential equation for $Q$ which contains both $Q^2$ and $\mathrm{d} Q/\mathrm{d} z$, with boundary conditions at the bottom and free surface, respectively. The eigenvalue $c$ is thence found using a shooting method. The fact that the system is nonlinear is a disadvantage which increases numerical cost and makes parallelization more cumbersome. We argue in Section \ref{sec:comparison} that our new method has several advantages over that of  \cite{dong12}, perhaps the greatest of which that our new method solves a linear second order differential equation .

\subsection{Analytical approximations}\label{sec:analapprox}

An altogether different approach is to find an explicit analytical expression dependent on $\mathbf{k}$ and $\mathbf{U}(z)$ which approximates $c(\mathbf{k})$. The most used of these was first presented by \cite{skop87} generalizing~\cite{stewart74}, and was developed to second order accuracy by \cite{kirby89}. This relation (generalized to the 3D case of a turning $\mathbf{U}(z)$) we call the 3DKC, and to leading order may be written (see \cite{ellingsen17})
\begin{equation} \label{KC}
\tilde{c}(\mathbf{k})\approx c_0(1-\delta); ~~ \delta = \int_{-h}^0\mathrm{d} z\frac{\mathbf{k} \cdot \mathbf{U}'(z)\sinh 2k(z+h)}{kc_0 \sinh 2kh}.
\end{equation}
Here, and for later reference, we define
\begin{align*}
\mathbf{U}_0 = \mathbf{U}(0), ~\mathbf{U}'_0 = \mathbf{U}'(0), ~\tilde{c}(\mathbf{k}) = c(\mathbf{k})-\mathbf{k}\cdot\mathbf{U}_0/k, 
\Delta \mathbf{U} =  \mathbf{U}-\mathbf{U}_0,~w_0=w(\mathbf{k},0)
\end{align*}
and $c_0 = \sqrt{(g/k+T k/\rho)\tanh kh}$. We recently proposed an alternative approximation
\begin{equation} \label{EL}
\tilde{c}(\mathbf{k}) \approx c_0(\sqrt{\delta^2+1}-\delta),
\end{equation}
which has certain advantages \cite{ellingsen17}. Both approximations come with a second order accurate extension providing excellent accuracy at far greater cost. 

For typical shear profiles occurring in oceanographic settings, both leading order approximations estimate $\tilde{c}(\mathbf{k})$ to within $5\%$ for all $\mathbf{k}$, often significantly better. For many practical cases this is quite adequate. However, for strongly sheared flows occurring in other systems, such as fast discharge of a surface jet into quiescent water or fast flow in a thin film, both approximations may become inaccurate and approximation \eqref{KC} could even become unphysical \citep{ellingsen17}. In a setting where computing cost is important, \eqref{KC} and \eqref{EL} would in practice be used ``blindly'' without any estimation of error, since this would in essence require a far more expensive second order calculation. 
We show in later sections that our method has an explicit error estimate as a built-in feature. It is moreover more robust than analytical approximations and with some extra iterations is able to tackle even profiles with extremely large shear and curvature where  \eqref{KC} and \eqref{EL} are poor. For moderately sheared profiles our method is comparable to \eqref{KC} and \eqref{EL} also in terms of cost, as discussed in Sec.~\ref{sec:analcomp}.

\section{Direct integration method}\label{sec:DIM}
We now consider the more general situation where water depth is allowed to vary, 
different from the existing approaches reviewed in Sec.~\ref{sec:appr}. We define 
\begin{equation}
[\bar{u},\bar{v},\bar{w},\bar{p},\bar{\zeta}](\mathbf{k},z)=[u,v,w,p/\rho,\zeta](\mathbf{k},z)/w(\mathbf{k},\eta^{(0)}).
\end{equation}
Here $u$ and $v$ are amplitudes of the horizontal velocities, $ p(z) $ is amplitude of the dynamic pressure, and $\zeta$ is the surface elevation, all in $\mathbf{k}$ space. 
The boundary value problem \eqref{eq:govn} may then be written 
following \citet{ellingsen17} and \citet{li18} ,
\begin{subequations}\label{eqset}
\begin{align} 
\bar{w}''-k^2\bar{w} ~ =& ~ \dfrac{\mathbf{k}\cdot\mathbf{U}''}{\mathbf{k}\cdot \Delta \mathbf{U}-k\tilde{c}}\bar{w}; ~~ -h<z<\eta^{(0)}; 
~~ \bar{w}(-h)=0; ~~ \bar{w}(\eta^{(0)}) =  ~1. \label{eq:RHE} \\
D_R(\mathbf{k}, \tilde{c})
&=0,\label{eq:dsp_R}
\end{align}
\end{subequations}
where
\begin{align} 
 D_R(\tilde{c})
&
\equiv  \tilde{c}^2 + \tilde{c} I_c(\tilde{c})-c^2_{0\bar{h}},
\label{eq:DR}
\\
I_c(\tilde{c})    & = \dfrac{\mathbf{k}\cdot\mathbf{U}'_0\tanh k\bar{h}}{k^2}+\tilde{c} \int\limits_{-h}^{\eta^{(0)}}\mathrm{d} z \dfrac{\mathbf{k}\cdot\mathbf{U}'' \bar{w}(\mathbf{k},z) \sinh k(z+h)}{k(\mathbf{k}\cdot \Delta \mathbf{U}-k\tilde{c})\cosh k\bar{h}},\label{eq:ic}\\
c ^2_{0\bar{h}} & =  \left( g/k+{Tk}/{\rho}\right) \tanh k\bar{h},\\
\bar{h}  & = h+\eta^{(0)}. 
\end{align} 
$ I_c $ denotes the contribution from a shear current on wave motion. For the case where $\mathbf{k}\cdot \Delta \mathbf{U}-k\tilde{c}=0$ for some critical depth $z_s\in \langle-h,\eta^{(0)}\rangle$, 
see discussion in section \ref{sec:crit}.
The implicit dispersion relation \eqref{eq:dsp_R} is found by multiplying \eqref{rayleigh} by $\sinh k(z+h)$, integrating with respect to $z$, and using the boundary condition \eqref{BC}
; refer to \citet{li18} for details. 

For later references, we define 
\begin{align}
 \tilde{\sigma}=& k\tilde{c}, ~\sigma_{0\bar{h}} = kc_{0\bar{h}},~\sigma(\mathbf{k},z) = k\tilde{c}-\mathbf{k}\cdot \Delta \mathbf{U}\equiv kc - \mathbf{k} \cdot \mathbf{U}.  
\end{align}

As is discussed in \cite{ellingsen17} for uniform water depth, when we further assume slow current compared to fast wave motion -- $ \mathcal{O}(U/c)\ll 1 $, \eqref{KC} is readily obtained from \eqref{eqset}; when the depth vertical shear is assumed to be small compared to wave motion, we obtain \eqref{EL}.    

The philosophy of the \textit{direct integration method} (DIM) is to treat Eqs.~\eqref{eqset} as two coupled equations with $ \bar{w} $ and $\tilde{c}$ as the unknowns and then to obtain the solutions of  Eqs.~\eqref{eqset} with numerical approaches.
Indeed, from a numerical point of view the DIM is simple to implement, combining only standard operations for solving linear inhomogeneous differential equations, numerical integration and root--finding, for which any of a number of methods may be used. 
We introduce an iterative scheme in Sec.~\ref{sec:num_Scheme} based on Newton's method; it is deliberately basic to ensure comparison with other methods be as fair as possible, and we do not claim it is the optimal choice.

\subsection{Error estimates} \label{sec:err}

Assume that $ \tilde{c}_\approx $ is an approximation to the exact solution $ \tilde{c}_e $ to \eqref{eq:dsp_R}. A Taylor expansion of \eqref{eq:dsp_R} about $ \tilde{c}=\tilde{c}_\approx $ reads (dependence on $\mathbf{k}$ is understood)
\begin{align}  \label{eq:tayD}
D_R (\tilde{c}_\approx+\Delta c)  = 
D_R (\tilde{c}_\approx) + \Delta c  \dfrac{\partial D_R }{\partial \tilde{c}}(\tilde{c}_\approx)
+
...
=0,
\end{align}
where 
$\Delta c = \tilde{c}_e-\tilde{c}_\approx$ 
and 
\begin{align}\label{dDdc}
\dfrac{\partial D_R }{\partial  \tilde{c}} = & 2\tilde{c}+ \left( I_c +\tilde{c} \frac{\partial I_c}{\partial\tilde{c}}(\tilde{c}) \right), \\
 \frac{\partial I_c}{\partial\tilde{c}}  = &  \int\limits_{-h}^{\eta^{(0)}}\mathrm{d} z \dfrac{\mathbf{k}\cdot\mathbf{U}''(\mathbf{k}\cdot \Delta \mathbf{U}) \bar{w}(\mathbf{k},z) \sinh k(z+h)}{k\sigma^2(\mathbf{k},z)\cosh k\bar{h}}. \label{dIcdc}
\end{align}

This yields an estimate of the relative error $R(\tilde{c}_\approx)$,
\begin{equation} \label{eq:rerr}
R(\tilde{c}_\approx) \equiv \left| \dfrac{\Delta c}{\tilde{c}_e}\right| 
\approx   
\left| \dfrac{D_R (\tilde{c}_\approx)}{\tilde{c}_\approx\frac{\partial D_R }{\partial \tilde{c}}(\tilde{c}_\approx)}\right|,
\end{equation}

Since $D_R$ and  $\partial D_R/\partial \tilde{c}$ need to be calculated anyway in order to solve \eqref{eq:RHE}, the error estimate \eqref{eq:rerr} can be calculated with very little additional cost.

\subsection{An explicit approximation of the DIM}
Based on \eqref{eq:rerr} , we obtain
\begin{equation}  \label{eq:tc_exp}
\tilde{c} \approx \tilde{c}_\approx - \dfrac{
D_R
(\tilde{c}_\approx)}{\partial
D_R
(\tilde{c}_\approx)/\partial \tilde{c} }
\end{equation}
which gives a good approximation of $ \tilde{c} $ if we insert either \eqref{KC} or \eqref{EL} as the $ \tilde{c}_\approx $, as was noted. This is a form of the solution with a single iteration 
. In principle, this returns the next order approximation of $ \tilde{c}_\approx $ to $ \tilde{c} $ and the accurate $ \bar{w} $  of the input $ \tilde{c}_\approx(\mathbf{k}) $. Compared to the 
second order corrections to 
\eqref{KC} \citep{kirby89} and \eqref{EL} \citep{ellingsen17}, 
\eqref{eq:tc_exp} 
offers much faster computation

%

\subsection{Critical layers} \label{sec:crit}

The integrand of the integral in \eqref{eq:ic} has poles whenever a critical layer 
exists, i.e.\ there exists a depth $z_c\in \langle-h,0\rangle$ so that $\mathbf{k}\cdot\mathbf{U}(z_c)=kc$.
This situation requires careful treatment of the integration path. 
In this circumstance one should in principle consider how the waves were created, treating the system physically as an initial value problem \citep{peregrine76}. One way to achieve this is to assume that a plane wave of frequency $\omega(\mathbf{k})$ has been created by a disturbance which has grown in time from zero at
$t=-\infty$,
in a manner proportional to  $ \mathrm{e}^{-\mathrm{i} \omega t + \mathrm{\epsilon} t} $ letting ($\mathrm{\epsilon}\to0^+$). 
Mathematically, this moves the integration path slightly off the real 
$z$ 
axis, making the integral in \eqref{eq:ic} well defined. Using the resulting integral, the real part $\tilde{c}$, pertaining to wave propagation, is kept in the limit $\epsilon\to 0^+$. (The imaginary part has a bearing on the stability of the critical layer; see e.g.\ discussion in \cite{velthuizen69, shrira93,ellingsen17}).

\subsection{Full flow field solution} \label{sec:gns}

A useful trait of the DIM is that in addition to the dispersion relation $c(\mathbf{k})$ 	it can calculate the full flow field with little extra cost. Given $c$ and $\bar{w}(z)$, remaining scalar flow fields $\bar{u},\bar{v},\bar{p},\bar{\zeta}$ are given by
\begin{subequations}\label{flowfield}
\begin{align}
\mathrm{i} k^2\bar{p}/\rho & = -(kc-\mathbf{k}\cdot\mathbf{U})\bar{w}'-\mathbf{k}\cdot\mathbf{U}'\bar{w}  , \\
k^2(kc-\mathbf{k}\cdot\mathbf{U}) \bar{u} & = \mathrm{i} k_x[\mathbf{k}\cdot\mathbf{U}' \bar{w} -(\mathbf{k}\cdot\mathbf{U}-kc) \bar{w}']-\mathrm{i} k^2 U'_x\bar{w},\\
k^2(kc-\mathbf{k}\cdot\mathbf{U}) \bar{v} & = \mathrm{i} k_y[\mathbf{k}\cdot\mathbf{U}' \bar{w} -(\mathbf{k}\cdot\mathbf{U}-kc)\bar{w}']-\mathrm{i} k^2  U'_y\bar{w},\\
k\tilde{c} \bar{\zeta} &= \mathrm{i} \bar{w}_0.
\end{align}
\end{subequations}
In order to obtain dimensional amplitudes we require the value of $w_0$. It can most often be readily calculated from the initial conditions of a particular problem. For a plane wave with known amplitude $a$, $w_0=-\mathrm{i} k\tilde{c} a$, following from the kinematic free-surface boundary condition.

\subsection{Group velocity from a kinematic approach}
Based on a kinematic approach, the group velocity can be readily obtained using the full derivatives on the relation
$ D_R(\mathbf{k},k\tilde{c})\equiv 0 $
 with respect to $ k_x $ and $ k_y $, i.e. 

\begin{align}
\nabla_\mathbf{k} D_R(\mathbf{k},\tilde{\sigma}) + \dfrac{\partial D_R(\mathbf{k},\tilde{\sigma}) }{\partial\tilde{\sigma} }  \nabla_\mathbf{k} (\tilde{\sigma})  \equiv  0,
\end{align}

in which the operator $ \nabla_\mathbf{k} = (\partial_{k_x} , \partial_{k_y})  $. The above relation further yields the group velocity defined 
\begin{align}
\mathbf{c}_g \equiv &  \nabla_\mathbf{k} (\tilde{\sigma}) +\mathbf{U}_0 = \dfrac{ \frac{\sigma_{0\bar{h}}}{\tilde{\sigma}}\mathbf{c}_{g_{\text{nc}}}- \frac{1}{2}\nabla_\mathbf{k} (kI_c) }{1+\frac{k I_c}{2\tilde{\sigma}}+\frac{1}{2}\frac{\partial I_c}{\partial\tilde{c}}(\tilde{c})} +\mathbf{U}_0,\label{cg_dimN}\\
\mathbf{c}_{g_{\text{nc}}} =&~ \dfrac{c_{0\bar{h}}}{2}\left(1+\dfrac{2k\bar{h}}{\sinh2k\bar{h}}  \right) \dfrac{\mathbf{k}}{k},
\end{align}
in which $\nabla_\mathbf{k} (kI_c)$ can be calculated via a numerical method and $\mathbf{c}_{g_{\text{nc}}} $ is the group velocity in the absence of a background flow. \eqref{cg_dimN} is applied to verify the group velocity described by \eqref{cg_dim}.

\subsection{Conservation equation of wave action} \label{sec:WAE}

We now proceed to the conservation equation of wave action (
henceforth called the wave action equation, WAE
)
  first developed by \cite{voronovich76} based on a dynamic approach and further developed by 
\citet{quinn17},
and show how the DIM can be used for a practical implementation of this equation. 
  The equation is in general applicable for any flow where there is large separation of lengthscales between the wave motion and vertical current variation on the one hand, and the horizontal variation of the current and bathymetry on the other, an approach well known from geometrical optics. This is a very typical situation in ocean and coastal modelling; see \cite{quinn17}.

The WAE can be derived by retaining terms to $ \mathcal{O}(\mathrm{\epsilon}^2) $ to obtain 

\begin{equation}  \label{eq:wae}
\dfrac{\partial I_{vs}}{\partial_t} +\nabla\cdot(\mathbf{c}_gI_{vs}) = 0. 
\end{equation}
in which 
\begin{subequations} \label{eq:cgvs}
\begin{align}
   I_{vs} = & \int\limits_{-h}^{\eta_0} \dfrac{\mathbf{k} \cdot \mathbf{U}''}{2\sigma^2k^2} w^2\mathrm{d} z +
       \left. \left[ \left(\dfrac{(g+Tk^2/\rho)}{\sigma^3}-\dfrac{\mathbf{k} \cdot \mathbf{U}'}{2\sigma^2k^2}\right) w^2\right]  \right|_{z=\eta^{0}} ,\\
   \mathbf{c}_gI_{vs} = & \int\limits_{-h}^{\eta^{(0)}} 
                             \left( 
                                       \dfrac{\mathbf{k} \cdot \mathbf{U}''}{2\sigma^2k^2} \mathbf{U}
                                      + \dfrac{\mathbf{U}''}{2\sigma k^2} 
                                       -\dfrac{\mathbf{k}}{k^2}
                             \right) w^2\mathrm{d} z \notag\\
                           & + \left. \left[ \left(  \left(
                                            \dfrac{(g+Tk^2/\rho)}{\sigma^3} -\dfrac{\mathbf{k} \cdot \mathbf{U}'}{2\sigma^2k^2}
                                     \right)\mathbf{U}- \dfrac{\mathbf{U}'}{2\sigma k^2} +\dfrac{(g+Tk^2/\rho)\mathbf{k}}{\sigma^2 k^2}
                                     \right) w^2 \right] \right|_{z=\eta^{(0)}} ,   
\end{align}
\end{subequations}
which are based on the notations defined herein and surface tension is considered and $ \eta^{(0)} \equiv 0$ for uniform water depth. 
$ w $ in \eqref{eq:cgvs} is the solution of the boundary value problem described by \eqref{rayleigh}. 
Refer to \cite{quinn17} for details and discussions. 

The DIM can be readily employed in \eqref{eq:cgvs}. As was noted, $ w = \bar{w} w_{\eta0} (\mathbf{k},\eta^{(0)}) $ in which $ \bar{w} $ is the solution of \eqref{eqset} and $ w_{\eta0} \equiv w(\mathbf{k}, \delta_tt,\eta^{(0)})$ is the amplitude of slow spatial and time variation, if any. We rewrite \eqref{eq:wae} and obtain 
\begin{equation}  \label{cg_dim}
\dfrac{\partial }{\partial_t} \left( \dfrac{w^2_{\eta0} }{k\tilde{\sigma}} \bar{I}_{vs}\right)  +\nabla\cdot\left( \mathbf{c}_g \bar{I}_{vs}\dfrac{w^2_{\eta0} }{k\tilde{\sigma}} \right) =0, 
\end{equation}
wherein
\begin{subequations}
\begin{align}
\bar{I}_{vs}= & \int\limits_{-h}^{\eta^{(0)}} \dfrac{\tilde{\sigma}}{\sigma } Q_{k}\bar{w}^2\mathrm{d} z + 
                \dfrac{\sigma^2_{0\bar{h}}}{\tilde{\sigma}^2\tanh k\bar{h} }
                -S_{k},\\
 \mathbf{c}_g =  & \dfrac{\tilde{\sigma}}{k}  \dfrac{\mathbf{N}}{\bar{I}_{vs}}  ,\\
 \mathbf{N}= &  \int\limits_{-h}^{\eta^{(0)}} 
     \left( 
           Q_{k} \dfrac{k\mathbf{U}}{\sigma } + \mathbf{Q} -\mathbf{k}
      \right) \bar{w}^2\mathrm{d} z 
       + \left. \left[  \left(
              \dfrac{\sigma^2_{0\bar{h}}}{\tilde{\sigma}^2 \tanh k\bar{h} } -S_{\mathbf{k}}
        \right)\dfrac{\mathbf{U}}{\tilde{c}}- \mathbf{S} +
        \dfrac{\sigma^2_{0\bar{h}}}{\tilde{\sigma}^2\tanh k\bar{h}  } \dfrac{\mathbf{k}}{k}
    \right] \right|_{z=\eta^{(0)}} ,\\
  \mathbf{S}= &\left. \dfrac{\mathbf{U}'}{2\tilde{\sigma} }\right|_{z=\eta^{(0)}} ,~S_{k} = \dfrac{\mathbf{k}\cdot\mathbf{S}}{k}, ~\mathbf{Q} = \dfrac{\mathbf{U}''}{2\sigma },~ Q_{k} = \dfrac{\mathbf{k}\cdot\mathbf{Q}}{k},
  \end{align}
  \end{subequations}
  where $ \mathbf{N}$, $ \bar{I}_{vs} $, $ \mathbf{S} $,  and $ S_k $ are dimensionless parameters, $ \mathbf{Q} $ and $\mathbf{Q_k}$ are of the same unit as $ k $. It is straightforward to find $ \mathbf{c}_g=
  \mathbf{c}
  _{g_{\text{nc}}} +\mathbf{U} $ when the current is irrotational, i.e.\ $ \mathbf{U}' = \mathbf{U}''\equiv0 $.

\section{Numerical scheme and example applications}\label{sec:num}

In this section we explore the numerical potentials of the DIM and demonstrate and test the DIM 
with an iterative algorithm for a range of different practical applications. Discussions regarding criterion of convergence and an estimate rate of convergence of the iterative algorithm are presented in Appendix \ref{sec:numDem}. 

\subsection{Iterative algorithm} \label{sec:num_Scheme}
Indeed, Eq.~\eqref{eq:dsp_R} is but a nonlinear scalar equation for $\tilde{c}$, as can be seen by noting that for some value of $\tilde{c}$ Eq.~\eqref{eq:RHE} determines $\bar{w}$ completely. Naturally a solution for $\tilde{c}$, and consequently $\bar{w}$, is iterative. The essence of the iterative algorithm is to consider $ \bar{w} $ as a function of an approximation  
$ \tilde{c}_{\approx} $
of $ \tilde{c} $
defined implicitly by Eq.~\eqref{eq:RHE} and evaluate $ \bar{w}( \tilde{c}_{\approx}) $ 
with an explicit scheme, as described below in step 1.

We discretize the $z$ axis into $N$  points $z_i,~i = 1,2,...,N$. The influence of $\mathbf{U}$ on the surface wave falls off for increasing $|z|$ as $\sim \exp(kz)$; For short waves compared to $h$, therefore, contributions from the water column will be negligible for $z$ below some threshold which depends on the desired accuracy. We choose the grid points equidistantly so that 
$z_0=\eta^{(0)}$
and 
$
kz_N = -\max[\alpha+2\ln(N/N_0), kh] ;
$ 
we found $\alpha= 3.5$ and $N_0=7$ to be suitable.  The ``bottom'' condition used is now $\bar{w}(z_N)=0$. 
This discretization works well for all cases we have tested 
yet we do not claim that it is the universally optimal choice.

Each iteration consists of three basic steps. 
\begin{enumerate}
	\item $\bar{w}({\tilde{c}}
	;z_i)$ is calculated from Eq.~\eqref{eq:RHE} using the value of $\tilde{c}$ from the previous iteration. For the first iteration an initial guess for $\tilde{c}$ is required. 
	\item An improved estimate of $\tilde{c}$ is calculated from Eq.~\eqref{eq:dsp_R} using $\bar{w}$ from step 1.	
	\item The error of the calculated $\tilde{c}$ is then estimated as described in section \ref{sec:err}, and iteration is terminated once a tolerance is reached. 
\end{enumerate} 

The number of iterations needed depends on the accuracy of the initial guess, yet in cases of moderate shear we shall see
that a single iteration with $c_0(k)$ as a naive initial guess, is accurate to within a few percent, sufficient for many purposes. 
A more accurate but more expensive first guess could be made using e.g.\ \eqref{KC} or \eqref{EL}. 

A number of standard methods exist for performing the first two steps of each iteration, the choice of which determines the computational cost and accuracy of the scheme of the DIM. In our implementation we have solved Eq.~\eqref{eq:RHE} [step 1] using a finite difference scheme with a 2nd order central difference approximation for $\bar{w}''$, resulting in a tridiagonal linear problem. Updating the value of $\tilde{c}$ using \eqref{eq:dsp_R} [step 2] we do with a single iteration of Newton's method,
i.e. 
	\begin{equation}
	\tilde{c}^{j+1} = \tilde{c}^{j} - \dfrac{D_R(\tilde{c}^{j})}{\partial D_R/\partial \tilde{c} |_{\tilde{c} = \tilde{c}^{j}}},
	\end{equation}
	in which $j$ denotes the $j^{\text{th}}$ iteration. 

All integrals with respect to $z$ 
we calculate using Simpson's method, all on the same grid of $z$-values. 

\subsubsection{Computational complexity}

The computational complexity of the above three steps is primarily from Step 1 and is of $ \mathcal{O}{(N)} $. If we implement `blind' predictions as \eqref{KC} or \eqref{EL} do, then one or two iterations (i.e. $ M=1 $ or $ M=2 $ in which $M$ is the total number of iterations) is sufficient to achieve the same accuracy as \eqref{KC} and \eqref{EL}, and an error estimate and a good approximation of $ \bar{w} $ in all evaluation points along the z-axis are also obtained at little or no extra cost,
see Sec.~\ref{sec:strongS}.

Furthermore, if we use either \eqref{KC} or \eqref{EL}  as an initial input of the DIM, The accuracy of the DIM from a single iteration is much increased. In most of the tested cases in the present paper, we use $ c_{o\bar{h}} $ as a naive guess to make comparisons of methods as fair as possible. 
Also calculating group velocity $ \mathbf{c}_g $ using Eq.~\eqref{cg_dim} incurs no additional complexity, and the overall complexity remains $\mathcal{O}(N)$.

\citet{quinn17} comment that "\textit{the WAE ... 
is difficult to apply to operational wave models as it is too computationally intensive: it is required to solve the Rayleigh equation for every node, frequency, direction, etc. at every time step}" for which reason they derive an approximate (explicit) form of \eqref{eq:cgvs} that suffers from a loss of accuracy. However, the DIM can change this picture radically; 
good 
accuracy may be achieved with  
the same complexity as numerical evaluation of the explicit equation of \citet{quinn17}.
As we will demonstrate, $ N=7 $ is enough for accuracy better than a few percent in typical cases and an example implementation of the DIM for slowly varying water depth is provided in Sec.~\ref{sec:WAEnum}.

\subsection{Turning profiles}\label{sec:turning}

\begin{figure}
	\graphicspath{{figures/}}
	\centering{\includegraphics[width=1\textwidth]{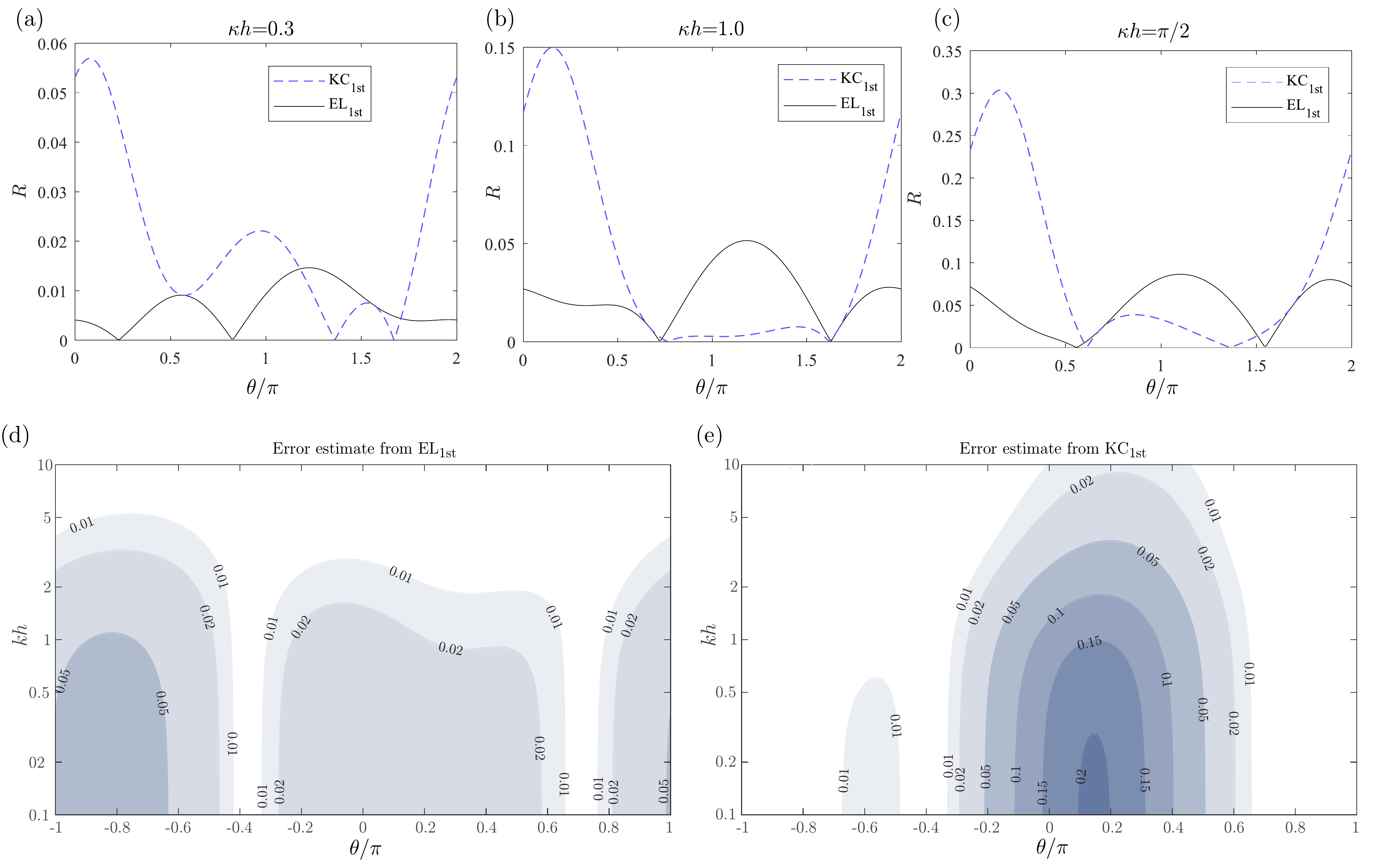}}	
	\caption{
		Error calculated for estimates from analytical approximations \eqref{KC} and \eqref{EL} (Denoted $\mathrm{KC}_\text{1st}$ and $\mathrm{EL}_\text{1st}$, respectively, for the turning profile \eqref{turning} with different parameters as shown. 
		In all panels $\mathrm{Fr}=U_0/\sqrt{gh}=0.5$ and $\alpha h=1$. In (a,b,c), $kh=1$, and in (d,e) $ \kappa h =1. $
	}
	\label{fig:turning}
\end{figure} 

In \cite{ellingsen17} approximations \eqref{KC} and \eqref{EL} were compared for a spiraling velocity field, but we were not in a position to compare the predictions to the accurate result since this was beyond the capabilities of the PLA, the best numerical method available at the time. With the DIM this task is easily accomplished. 

As in \cite{ellingsen17} we consider the profile, 
\begin{equation} \label{turning}
\mathbf{U}(z) =U_0\sinh \alpha(z+h)(\mathbf{e}_x \cos \kappa z + \mathbf{e}_y\sin\kappa z).
\end{equation}
Choosing $\alpha h=1$ corresponding to a wavelength $2\pi h$, we consider waves 
propagating in different directions $\theta$.

As conjectured in \cite{ellingsen17} the 3DKC estimate \eqref{KC} performs relatively poorly for the turning profiles compared to unidirectional examples, particularly in the area
$0.1\lesssim\theta/\pi\lesssim 0.2$ 
in this example. 
As shown in Fig.~\ref{fig:turning}a-c, where $kh=1$, 
this holds true even for very weakly turning profiles with $\kappa h =0.3 $ and $\kappa h=1$, and worsens with stronger changes of direction. 
The 3DKC tends to perform well in the vicinity of $\theta=\pi$, yet the approximation \eqref{EL} appears to be more consistent. 
The error estimates for the two analytical approximations as functions of $\theta$ and $kh$ are shown in Fig.~\ref{fig:turning}d,e. For both cases performance is least good for long wavelengths. 
While one should not draw too strong conclusions based on a single example, Fig.~\ref{fig:turning} seems to indicate that approximation \eqref{EL} is preferable for turning profiles. A more careful analysis is beyond the scope of the present Article. 

\subsection{Strongly sheared profiles} \label{sec:strongS}
\begin{figure}
	\graphicspath{{figures/}}
	\centering{\includegraphics[width=1\textwidth]{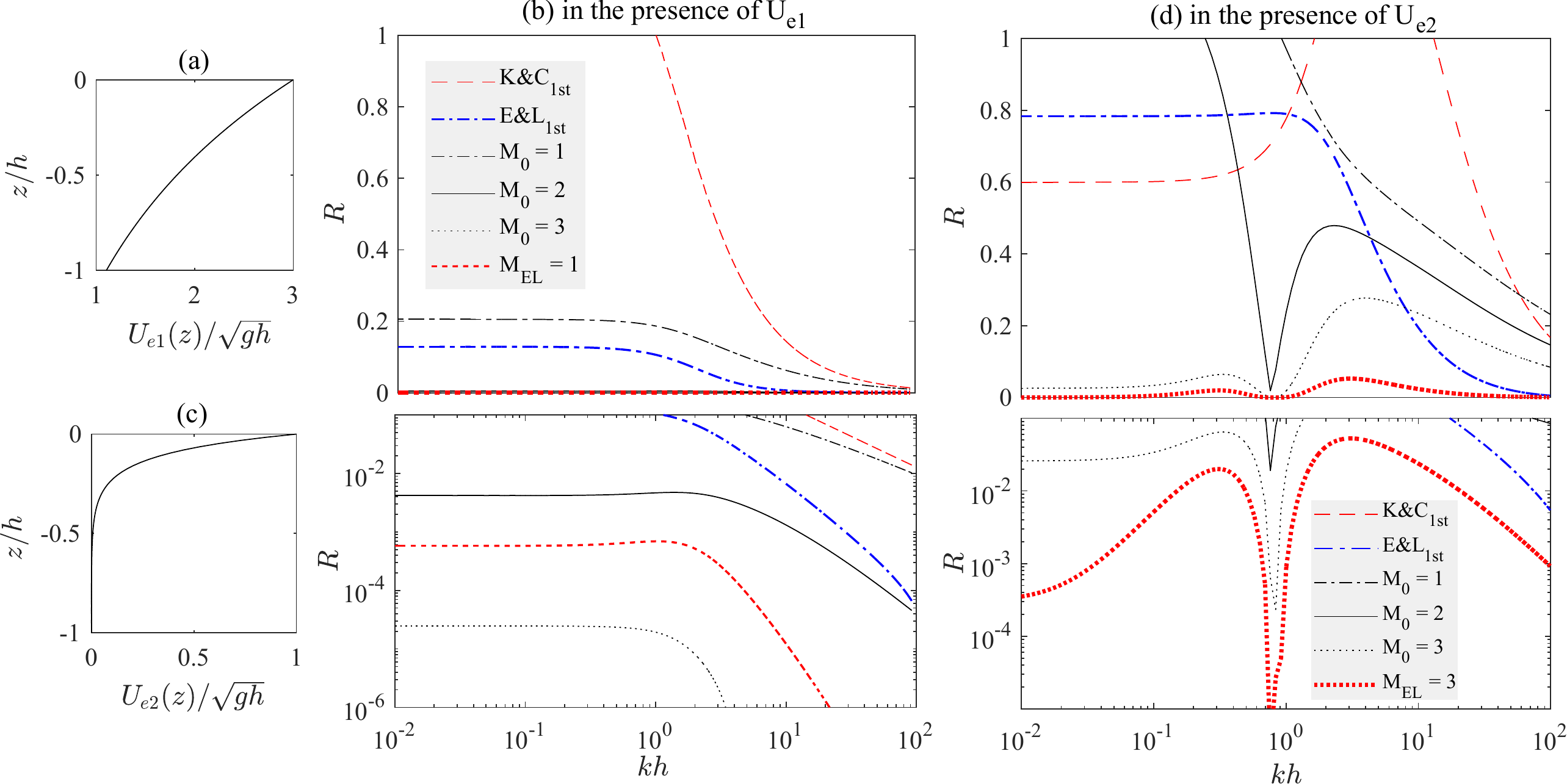}}	
	\caption{
		Applying the DIM to two very strongly sheared velocity profiles given in (a,b): Eq.~\eqref{profile1} and (c,d): Eq.~\eqref{profile2}. Velocity profiles are shown in (a,c), corresponding to plots of relative error $R$, Eq.~\eqref{eq:rerr} for different calculations. $\mathrm{K\&C}_\text{1st}$: Approximation \eqref{KC}, $\mathrm{E\&L}_\text{1st}$: approximation \eqref{EL}, $M_0=1,2,3$: DIM with initial guess $c_0$ using $1,2$ and $3$ iterations, respectively, $M_\text{EL}=1,3$: DIM using \eqref{EL} as initial guess.  
	}
	\label{fig:extreme}
\end{figure} 

In this section we demonstrate how the DIM can readily handle very strongly sheared profiles, of which we consider two example flows 
\begin{subequations}\label{extreme} 
\begin{align}
U(z) &= 3\sqrt{gh}\exp(z/h), \label{profile1} \\
U(z) &= \sqrt{gh}\exp(10z/h).\label{profile2}
\end{align}
\end{subequations}
For both examples the first--order analytical approximation \eqref{KC} performs poorly, and even yields unphysical results for some wavelengths. Also approximation \eqref{EL} is poor for profile \eqref{profile2} (see further details and discussions in \citet{ellingsen17}). 

The profiles \eqref{extreme} are too strongly sheared to represent oceanographic flows, but could realistically occur in other flow settings. Eq~\eqref{profile1} could represent, e.g., for a flow of surface velocity $4$m/s of $40$cm depth, for example over a local shallow in a river, or a film flow of $1$cm depth with surface velocity $60$cm/s, which is readily produced. Eq~\eqref{profile2} might represent a surface jet due to discharge of a fast flow into a still water reservoir, e.g.\ a jet speed of $3$m/s over $1$m depth. 
An oceanographic situation where shear can be strong is a relatively short period after the onset of wind over the surface \cite{caulliez98}.

Relative errors for calculations of phase velocity for waves propagating along the direction of the flow as a function of wave number are shown in Fig.~\ref{fig:extreme}. It is clear to see that DIM converges quickly, producing accurate results in these cases with only $3$ iterations even with a poor initial guess $c_0$.
For the velocity profile of Fig.~\ref{fig:extreme}a, approximation \eqref{EL} is reasonably successful (within 20 \% of the true value) and using this as initial guess the accuracy is better than $0.1\%$ with only one iteration of the DIM. 

In a sense, a single iteration of DIM can be interpreted as sibling of the analytical approximations, in that it also constitutes a single integration of a functional of $U(z)$ along the $z$ axis. Just like \eqref{KC} and \eqref{EL} this may be inaccurate for extreme profiles. Unlike these, however, the DIM can simply be iterated whereas the second order extensions of \eqref{KC} and \eqref{EL} are far more expensive. 

One should note that when the initial guess is very poor, as for example when using $c_0$ in the examples \eqref{extreme}, the error estimate \eqref{eq:rerr} is far from its real value. Nevertheless it will produce an estimate that is higher than any realistic error tolerance, ensuring that more iterations are performed, with correspondingly more accurate error estimates. 

\subsection{Velocity field}	 
Using Eqs.~\eqref{flowfield}, the full flow field is readily obtained by the DIM at little extra cost. As demonstration we calculate the wave--induced velocity field due to an initial surface perturbation in 2 dimensions, on a shear flow with the profile
\begin{equation}  \label{eq:Exp}
U(z) =  0.1 \sqrt{gh} (\exp{(6z/h)}-1) 
\end{equation}
representative of a surface drift layer. 
To our knowledge such a Cauchy--Poisson problem has not been considered before for the velocity field in the presence of a velocity field of non--uniform vorticity.

\begin{figure*}
	\graphicspath{{figures/}}
	\includegraphics[width=1\textwidth]{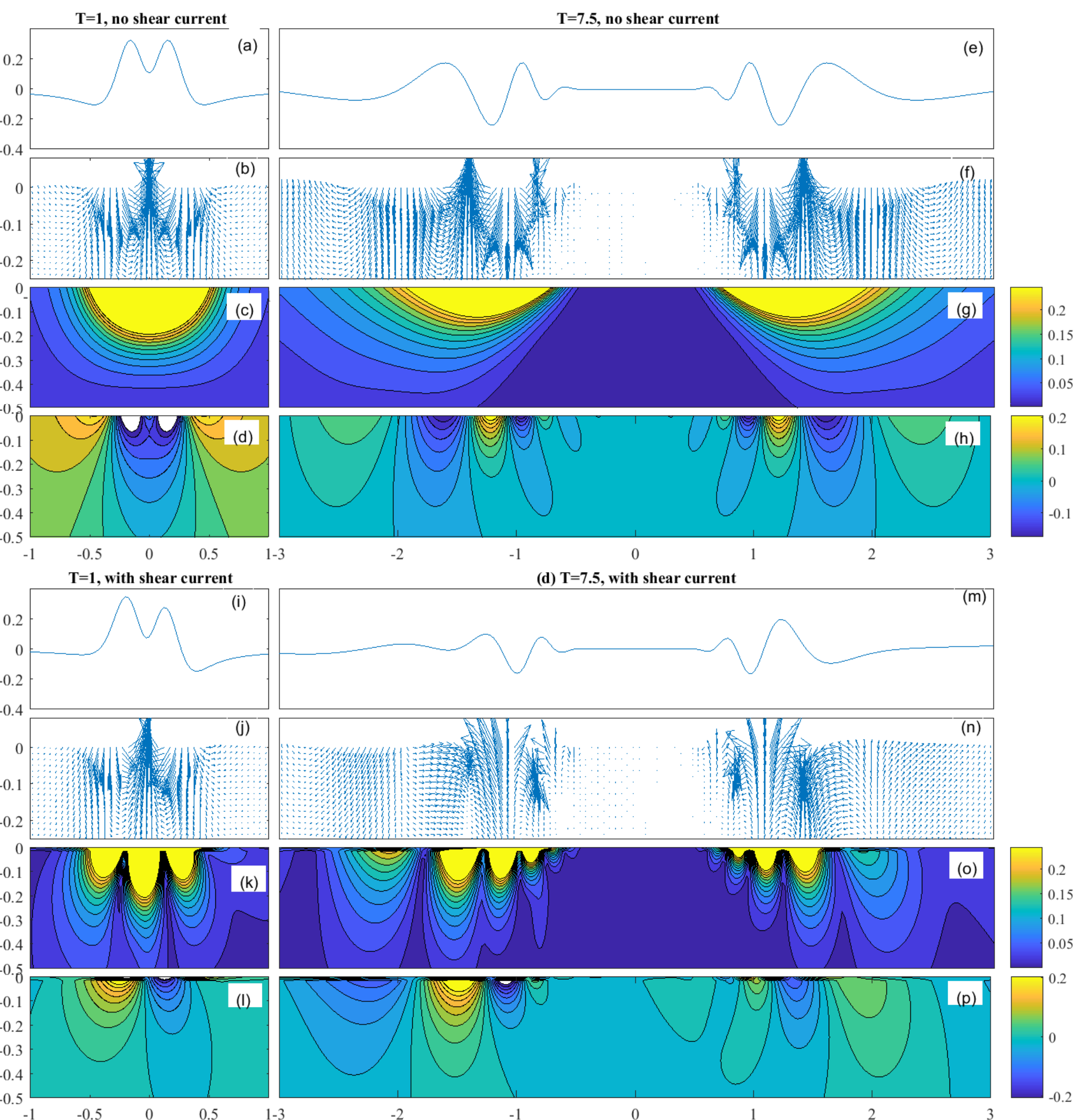}
	\caption{
		Waves from initial pressure pulse with (i-p) and without (a-h)  shear current profile \eqref{eq:Exp} at times $T=1$ (left column) and $7.5$ (right column) . Surface elevation (a,e,i,m), velocity field (b,f,j,n), velocity magnitude (c,g,k,o) and pressure (d,h,l,p). Video in supplementary materials.
	}
	\label{fig:velocity}
\end{figure*}

Fig.\ref{fig:velocity} depicts the velocity field and surface elevation in 2D at two instant times $ T=t\sqrt{g/h} = 1 $ and $5$ generated by an initial impulsive pressure $ \dfrac{\hat{p}_{\text{ext} }}{\rho g h}= \exp(-\dfrac{\pi^24x^2}{h^2})  \delta(t) $. Results in the presence and absence of the sub-surface velocity profile \eqref{eq:Exp} are shown in the right and left columns of Fig.\ref{fig:velocity}, respectively. The velocities and pressures plotted are the perturbation fields, i.e., after subtracting their values when no waves are present. 

As one would expect in light of previous studies (e.g. \citet{ellingsen14, li17a}) the surface shape is changed visibly, yet moderately by the shear flow. The velocity and pressure fields, on the other hand, are strikingly different in qualitative appearence. Without shear current the velocity magnitude beneath the surface elevation has slow spatial variation with only the direction changing rapidly. Not so in the presence of the sheared current, in which case there are several highly distinct regions directly beneath the largest surface excitations with far lower absolute velocities. In the present example these regions are near--vertical in shape. The rotating wave motion undergoes a depth--dependent phase shift due to the depth--varying velocity field. 

While only a single example, these observations seem to indicate that the velocity field beneath waves can be strongly affected by e.g.\ a wind--driven shear layer, as \eqref{eq:Exp} might represent. This is a potentially important observation, since the near--surface fluid mechanics of the oceans is crucial for processes in oceanography and climate modelling, in particular transportation of nutrients and algae, and mixing of warmer and colder waters. The effect of shear--modified wave motion on sub-surface turbulence intensity, Reynolds stresses and thermal mixing are all virtually unknown and make for an important as well as intrinsically interesting area of future study. We have demonstrated how the DIM can offer a fast and computationally cheap first insight.

\subsection{Wave amplitudes over slowly varying water depth } \label{sec:WAEnum}
\begin{figure*}
	\graphicspath{{figures/}}
	\centering
	\includegraphics[width=0.8\textwidth]{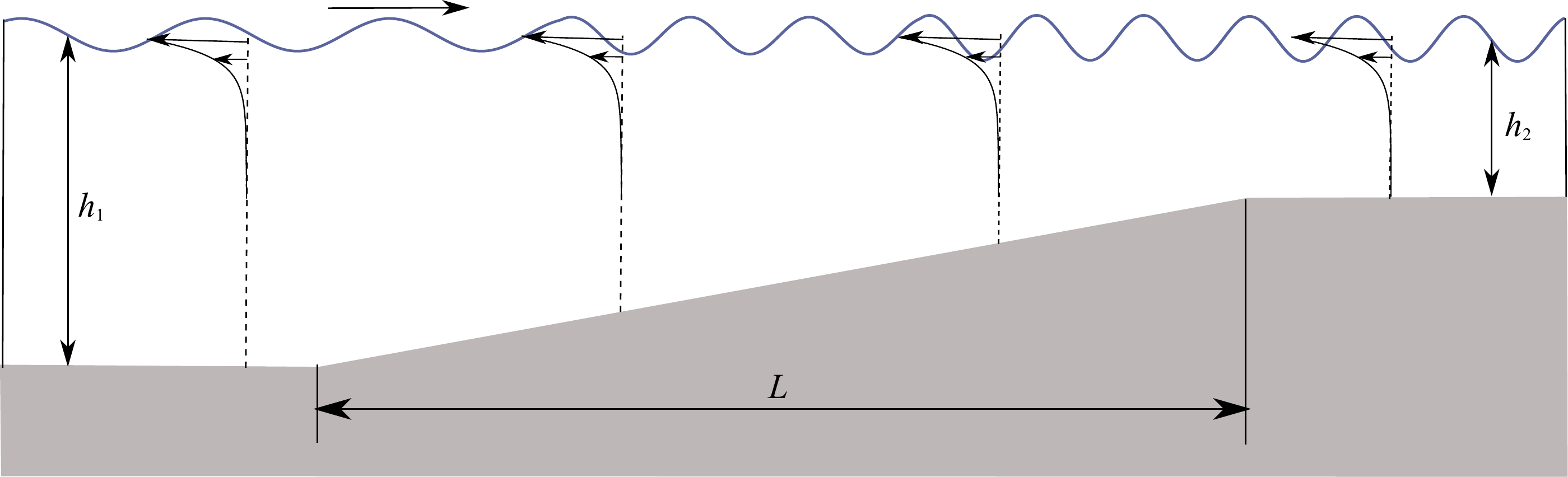}
	\caption{
		Waves propagating over slowly  varying water depth in the presence of a wind-induced current.  In the figure, $ h_1 $ and $ h_2 $ are water depth and $ L $ is the characteristic length in the horizontal plane. The slowness of depth variation is indicated by $ |h_1-h_2|/L\ll 1 $ . 
	}
	\label{fig:slope}
\end{figure*}	

As noted in section \ref{sec:WAE}, the DIM calculates wave rays and amplitudes with computational complexity of order $ \mathcal{O}(N) $. We now consider as an example, steady wave propagation over a sloping seabed in the presence of a constant wind--induced shear current, as depicted in Fig.~\ref{fig:slope}. Due to waves being steady, the time--dependent term in the wave action equation is zero, hence
\begin{equation}  \label{eq:steadycg}
\partial_x \left( c_{gx} \bar{I}_{vs}\dfrac{w^2_{\eta0} }{\tilde{\sigma}} \right) +  
\partial_y \left( c_{gy} \bar{I}_{vs}\dfrac{w^2_{\eta0} }{\tilde{\sigma}} \right) =0, 
\end{equation}
which can be solved numerically using ray theory when information at a point is specified. For more details in the absence of a shear current, one may refer to \citet{mei89}. 

As depicted in Fig.~\ref{fig:slope}, we consider a linearly varying seabed $ h(x) $ that has a negligible effect on the wind-induced current, implying $ \eta^{(0)} = 0 $ and $ |h_1-h_2|/L\ll \mathrm{\epsilon} $ where $ L $ is the characteristic horizontal length; A surface drift is expressed $ U = \mathrm{Fr}_h\sqrt{gh_2} (\exp(6z/h_2)-1) $; $ H = h_1/h_2>1$ is defined where $ h(x_1) = h_1 $ and $ h(x_2)=h_2 $ 
	and $\mathrm{Fr}_h = U_0/\sqrt{gh_2} $.
 Hence, \eqref{eq:steadycg} yields
\begin{equation} \label{eq:amp}
\dfrac{A_2}{A_1} = \sqrt{\dfrac{c_{gx}(x_1)}{c_{gx}(x_2)}}. 
\end{equation}
where $ A_2 =  A(x_2)$ and $ A_1 = A(x_1) $. For cases with the absence of a shear current, we use the subscript `nvs' to note. Eq.~\eqref{eq:amp} is solved using the DIM wherein the frequency of an incident wave remains constant over the varying water depth. 

\begin{figure*}
	\graphicspath{{figures/}}
	\centering
	\includegraphics[width=1\textwidth]{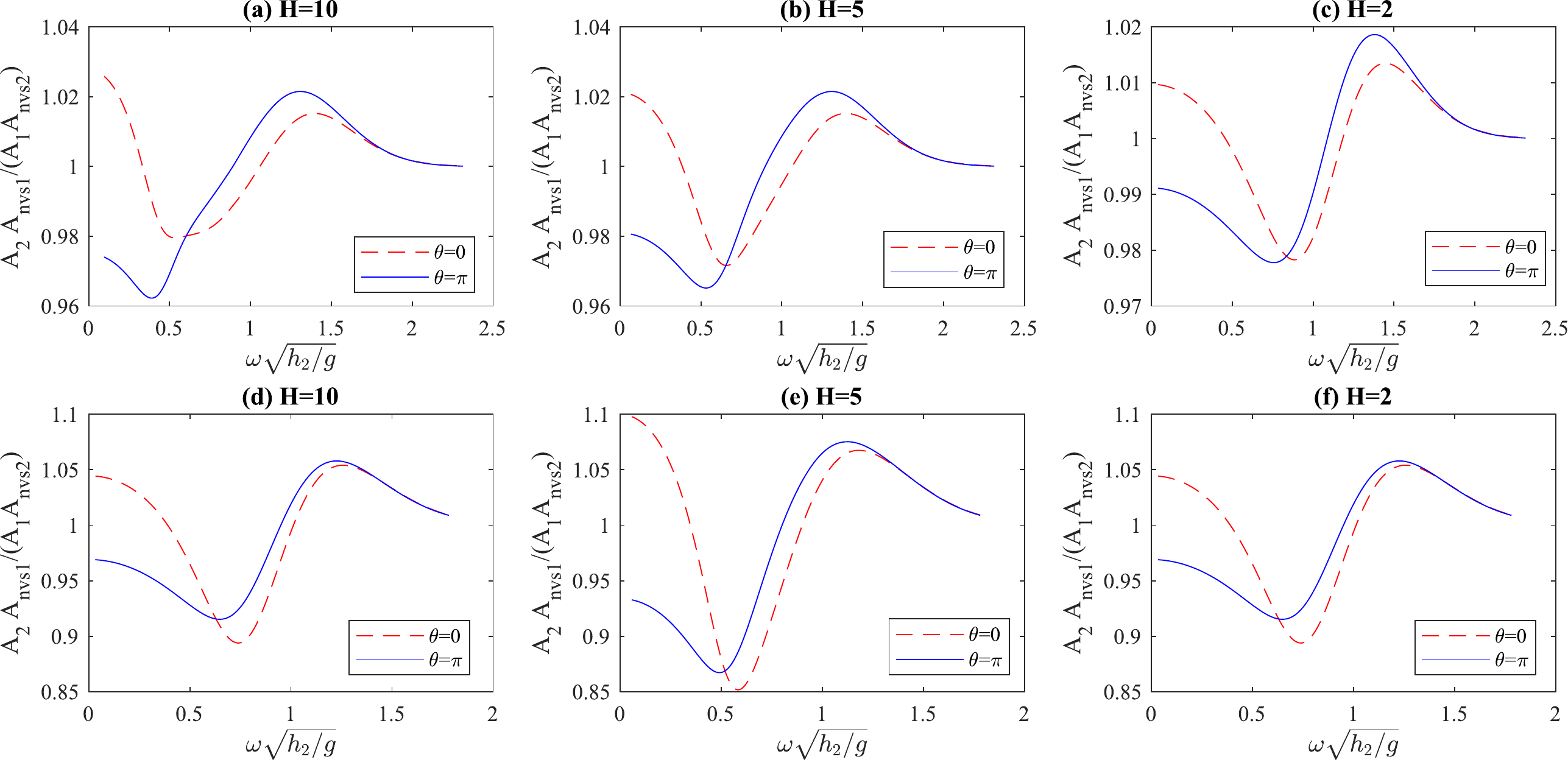}
	\caption{
		Comparison of wave amplitudes in the presence and absence of a vertical shear flow at different relative water depth $ H=h_1/h_2 $. In the figure, (a-c) $ \mathrm{Fr}_h = 0.1 $; (d-f) $ \mathrm{Fr}_h = 0.4 $. 
	}
	\label{fig:amplitude}
\end{figure*}

Fig.~\ref{fig:amplitude} compares the amplitude change of waves at different frequencies with and without the shear when $ \mathrm{Fr}_h=0.1 $ and $ \mathrm{Fr}_h=0.4 $, respectively. It is seen the amplitude change of a wave oscillates at a specific frequency can be over(under) estimated by up to $ 6 \% $ ($ \mathrm{Fr}_h=0.1 $ ) or to $ 15 \% $ ($ \mathrm{Fr}_h=0.4$ )  when a subsurface shear is neglected. This is in keeping with the conclusions drawn by  \citet{zippel17}. The DIM offers a viable route to this end.

\section{Comparison with other approaches} \label{sec:comparison}

In this section we will compare the DIM with the iterative scheme to existing approaches for calculating dispersion relation $c(\mathbf{k})$, both numerical methods with arbitrary accuracy, and analytical approximations with theoretical error. These two categories of existing approaches were reviewed in Sec.~\ref{sec:arbacc} and \ref{sec:analapprox}, respectively.

\subsection{The piecewise--linear approximation} 
We compare first with the piecewise--linear approximation (PLA, c.f.\ Sec.~\ref{sec:arbacc}). 
Having implemented both the PLA (albeit only for unidirectional flow; see \cite{smeltzer17} for full details) and the DIM, we are in a position to directly compare the two methods. While the choice of method will always remain a point of preference of the user, we find it hard to imagine a practical application for which the DIM is not preferable to the PLA.

The one point in favor of the PLA compared to the DIM is its physical transparency. It is very easy to follow all physical quantities explicitly throughout calculations. The rate of convergence of the PLA is typically similar to that of the iterative approach of the DIM employed here, $\sim N^{-2}$ (see \cite{zhang05,smeltzer17} and Fig.~\ref{fig:pg}). 
The DIM reaches a higher convergence rate if a numerical approach of higher accuracy (than the second-order central difference approximation for $ \bar{w}'' $ ) is used. Both the DIM and PLA can approximate the full flow field with little extra effort. 

However, from a numerical point of view the DIM has a number of advantages. Firstly its implementation is significantly simpler. The PLA initially produces $N+1$ solution to the dispersion relation, $N-1$ of which are artifacts of the abrupt change of vorticity at layer interfaces. There are reliable ways to deal with this problem \citep[see ][]{smeltzer17}, but it remains a hurdle. Secondly, and for the same reason, computation is less costly, since detecting the correct eigenvalues of $\tilde{c}$ can be the most costly part of the PLA calculation. Thirdly, the DIM can easily handle cases where $\mathbf{U}(z)$ changes direction with depth, as demonstrated in Sec.~\ref{sec:turning}. Short of a more sophisticated PLA implementation than has been developed to date to our knowledge, this would double the number of free coefficients to be determined in the PLA. Fourthly, the DIM comes with a direct estimate of the error with very little extra cost; error estimation using PLA is not straightforward (the obvious solution is by comparing results from different values of $N$; however we cannot preclude that a more intelligent way can be found).

\subsection{Dong \& Kirby's method}  

Also compared to the method of \cite{dong12} the DIM definite advantages. 
Foremost of the advantages of the DIM over Dong \& Kirby's (DK) procedure is that the ordinary differential equation to be solved, \eqref{eq:RHE}, is linear, allowing a fast and cheap solution e.g.\ with a finite difference scheme. Dong \& Kirby's method (DK), on the other hand solves a nonlinear ordinary differential equation of the function $Q(z)=w(z)/w'(z)$, determining the eigenvalue $c$ with a shooting method. For the DIM the linear solution is easily parallelizeable, able to perform all operations for an array of $\mathbf{k}$ values at once. Should the full flow field be required, the DIM produces this automatically whereas DK produces only $Q(z)$, requiring further integration in order to obtain the velocity via $w(z) = \exp(\int \frac{\mathrm{d} z}{Q})$. The explicit error estimate of DIM could be a further advantage, although convergence of DK's shooting scheme can also be used for error estimation. Also compared to the DK procedure it is our opinion that the DIM is superior for all purposes we can think of.

\subsection{Analytical approximations}\label{sec:analcomp}

We finally compare the DIM to analytical approximations with theoretical error presented in Sec.~\ref{sec:analapprox}. 
Obviously such a comparison will be context dependent, since the philosophies behind numerical and analytical approximations are fundamentally different: 
A numerical implementation of the DIM has arbitrary accuracy 
while approximations \eqref{KC} and \eqref{EL} have finite theoretical error no matter the accuracy with which the associated integrals are evaluated. With this in mind, we have tried to make comparison as fair as possible. We will presume a context which involves calculating $c(\mathbf{k})$ for a large range of $\mathbf{k}$ spanning all directions and several orders of magnitude in terms of wavelengths  ranging from very deep to very shallow waves, and that at least a rough notion of the calculation error is desired. 

There are obvious advantages to using DIM rather than analytical approximations such as \eqref{KC} and \eqref{EL}, beyond the mere fact that arbitrary accuracy can be achieved. Firstly, as demonstrated in Fig.~\ref{fig:extreme}, DIM can easily handle difficult cases where analytical approximations perform poorly, without greatly increasing computational cost compared to weakly sheared flows. Secondly, the DIM yields the full velocity and pressure fields, whereas \eqref{KC} and \eqref{EL} provide $c(\mathbf{k})$ only. Perhaps most pertinently, DIM 
facilitates low-cost error estimation, whereas in a context where computational cost is of importance, first--order analytical approximations must in practice be used ``blindly'', without any control of the error made, since an error estimate will essentially involve going to far more costly second--order approximations (refer to \cite{kirby89,ellingsen17}). 

We choose a context where neither of these advantages play a role, and where analytical approximations \eqref{KC} and \eqref{EL} are routinely in use today, namely for quick estimation of dispersion relations as part of a bigger oceanographic or coastal flow simulation, c.f.\ e.g.~\cite{elias12,kumar12}. For this purpose the analytical approximations \eqref{KC} and \eqref{EL} are very suitable: shear profiles are typically not strongly sheared so that analytical approximations are typically well within accuracy requirements. The analytical approximations are also cheap to calculate compared to numerical schemes reviewed in Sec.~\ref{sec:arbacc}. For fairness, however, we will also employ the
{iterative DIM algorithm}
``blindly'', spending no time on error estimation and tolerance comparison. For the {iterative algorithm} this amounts to a cost reduction of only a few percent for the small number of iterations we consider. Integrals in analytical approximations as well as the {iterative algorithm} are calculated with the same 2nd order accurate method (Simpson's method), with the number of grid points as specified. For some $N$, the same discretization is appropriate for both schemes. For analytical approximations, the smallest chosen $N$ is just large enough so that calculation of the integral is not the main source of error (naturally this can only be checked \emph{a posteriori} with an expensive error calculation, hence a somewhat higher $N$ should be used in practice). In order not to favour any particular range of wavelengths we calculate values for a grid of $512\times 512$ values of $\mathbf{k}$ covering values $|\mathbf{k}|h$ from $10^{-2}$ to $10^2$ isotropically. The maximum value of $R$ from these values is presented. Since calculational times are essentially identical for \eqref{KC} and \eqref{EL}, and their theoretical errors are similar in magnitude for moderately sheared flows, we include results only for the 3DKC \eqref{KC} from \cite{skop87,kirby89}. 

Calculation times are given in Table \ref{tb:times} for the wind-driven profiles shown in Fig.~\ref{fig:wind}a. $c(\mathbf{k})$ was calculated 
using MATLAB 
for a grid of $512\times 512$ values of $\mathbf{k}$ on  a standard desktop computer (8 processors: Intel i7-4770 3.4 MHz, 32 GB RAM). Naturally computational cost depends on the choice of methods for calculation of integrals and solution of the boundary value problem \eqref{eq:RHE} as well as for the analytical approximations. Calculations were parallelized, calculating the full matrix of $\mathbf{k}$ values simultaneously. 

\begin{table}[] 
	\centering
	\caption{Computation times for calculating $c(\mathbf{k})$ for a grid of $\mathbf{k}$-values using the {iterative algorithm} and analytical approximation \eqref{KC}. See main text for further details.}
	\begin{tabular}{|l|l|l|l|l|l|l|l|l|}
		\hline
		&                                           &                                            & \multicolumn{2}{l|}{}                                                     & \multicolumn{4}{l|}{Max R}                                                     \\ \cline{6-9} 
		&                                           &                                            & \multicolumn{2}{l|}{\multirow{-2}{*}{Time, $ 512^2 $ values}} & \multicolumn{3}{l|}{$ c_0 $ init.}   & $ c_{kc} $ init.                                \\ \cline{4-9} 
		\multirow{-3}{*}{}        & \multirow{-3}{*}{N}                       & \multirow{-3}{*}{M}                        &$  c_0 $ init.                          & $ c_{kc} $ init.                         & profile 1 & profile 2 & profile 3 & profile 1                                  \\ \hline
		&                                           & 1                                          & 0.181                               & 0.241                               & 0.0242    & 0.0318    & 0.0154     & 6.50E-03                                   \\ \cline{3-9} 
		&                                           & 2                                          & 0.302                               & 0.361                               & 7.71E-04  & 4.51E-04  & 9.49E-05  & 3.04E-05                                   \\ \cline{3-9} 
		& \multirow{-3}{*}{7}                       & 3                                          & 0.426                               & 0.486                               & 4.06E-05  & 1.44E-05  & 1.36E-06  & 1.69E-06                                   \\ \cline{2-9} 
		&                                           & 1                                          & 0.520                               & 0.599                               & 0.0245    & 0.0318    & 0.0153    & 1.26E-03                                   \\ \cline{3-9} 
		&                                           & 2                                          & 0.893                               & 0.929                               & 8.58E-04  & 4.64E-04  & 1.07E-04  & 7.35E-05                                   \\ \cline{3-9} 
		& \multirow{-3}{*}{16}                      & 3                                          & 1.251                               & 1.308                               & 5.13E-05  & 1.51E-05  & 1.62E-06  & 4.56E-06                                   \\ \cline{2-9} 
		& \multicolumn{1}{c|}{}                     & 1                                          & 1.178                               & 1.232                               & 0.0244    & 0.0317    & 0.0153    & 1.00E-03                                   \\ \cline{3-9} 
		& \multicolumn{1}{c|}{}                     & 2                                          & 1.871                               & 1.951                               & 8.65E-04  & 4.68E-04  & 1.08E-04  & 5.94E-05                                   \\ \cline{3-9} 
		\multirow{-9}{*}{DIM}     & \multicolumn{1}{c|}{\multirow{-3}{*}{32}} & 3                                          & 2.576                               & 2.554                               & 5.23E-05  & 1.52E-05  & 1.65E-06  & 3.75E-06                                   \\ \hline
		& 7                                         &                   & \multicolumn{2}{l|}{0.075}                                                & 0.045     & 0.036     & 0.024     &                   \\ \cline{2-2} \cline{4-8}
		& 16                                        &                   & \multicolumn{2}{l|}{0.141}                                                & 0.023     & 0.033     & 0.011     &                   \\ \cline{2-2} \cline{4-8}
		\multirow{-3}{*}{KC$ _{1st} $} & 32                                        & \multirow{-3}{*}{---} & \multicolumn{2}{l|}{0.282}                                                & 0.022     & 0.031     & 0.010     & \multirow{-3}{*}{---------} \\ \hline
	\end{tabular}
\end{table}\label{tb:times}

A number of interesting observations can be made. Firstly, discretizing the $z$ axis with only $N=7$ points and running a single iteration is sufficient to achieve accuracy at the level of the theoretical error of the 3DKC even though the initial guess $\tilde{c}=c_0$ is naive and does not make use of any knowledge of $U(z)$. Using instead $c_\text{KC}$ as initial guess the error is reduced by a factor $ 10 $ or more, although calculation is then necessarily more expensive. 

Although results show that with $N=7$, 3DKC gives errors $<5\%$, likely to be adequate in many cases, the error in the integral evaluation is still a significant and uncontrolled contributor to the maximum error, thus without an error estimate a higher value of $N$ should be used in practice. Using $N=16$, the calculation time is only slightly lower than the 
$N=7$ DIM calculation which has essentially the same accuracy.
In contrast, increasing $N$ for the 
iterative DIM algorithm

does not significantly improve accuracy, which depends almost exclusively on the number of iterations in the examples shown. Based on this we opine that it is fair to say that the 
iterative DIM algorithm
can realistically compete with analytical approximations even in cases where the latter is particularly suitable and in routine use, and given the advantage of easy control of errors, can be a very viable alternative for implementation in oceanographic models such as detailed in \cite{elias12,kumar12}. 

Including an error estimate for the iterative DIM algorithm only has numerical cost in the last iteration because both integrals calculated to estimate $R$ in Eq.~\eqref{eq:rerr} are made use of in the next iteration if the latter is performed. Calculating the estimated $R$ then incurs approximately half the cost of the next, unevaluated, iteration. Checking the relative error for $N=7, M=1$, for example, increases calculation time to about $0.24$, an increase of less than $30\%$. Relative increase in cost is obviously smaller for higher $M$.  

Should higher accuracy be required, results in Table \ref{tb:times} also show that additional iterations are significantly cheaper than the first.

\section{Conclusions}\label{sec:concl}

We have developed a direct integration method (DIM) for linear surface waves travelling at arbitrary angles atop a horizontal background  current $\mathbf{U}(z)${allowing}
slowly varying barthymetry; both the magnitude and direction of the current may vary arbitrarily as a function of depth. 
{In particular, when depth is constant the DIM allows efficient evaluation of the dispersion relation over arbitrary shear. }
We also derive the full approximate flow field solution of the wave-shear current-sloping seabed system and revisit the conservation equation of wave action for which the DIM offers cost-efficient means of numerical evaluation.

We implement the DIM in an iterative procedure using standard constituent methods due to \citet{quinn17}.
The iterative DIM algorithm comes with a built--in error estimate for comparison with a tolerance level and can make the DIM somewhat explicit by limiting the total number of iterations with a reasonable initial guess.

We argue that the DIM is superior to existing calculation methods with arbitrary accuracy {with constant depth}
, namely the piecewise--linear approximation (PLA) in which the water column is divided into $N$ artificial layers with linear $\mathbf{U}(z)$ within each \citep{zhang05,smeltzer17}, and a shooting method due to  \cite{dong12} (DK).Compared to the PLA, the DIM is at least as fast at comparable accuracy, considerably easier to implement, and can easily handle turning profiles. The DK solves a non-linear differential equation and is considerably slower, and arguably numerically more complicated, than the DIM. 

Compared to analytical approximations such as those of \cite{skop87}/\cite{kirby89} (KC)  or \cite{ellingsen17} (EL),  
the DIM has some obvious advantages beyond the mere fact that arbitrary accuracy can be achieved; it  can easily handle difficult, strongly sheared flow situations where the above analytical approximations perform poorly; it yields the full flow field with little extra effort; and it  provides an estimate of the relative error of the intrinsic phase velocity at only slightly increased cost, whereas the analytical approximations must either be used without any control of errors, or a far more expensive 2nd order estimate must be calculated. 

The respective importance and relevance of the above advantages will naturally depend on the context in which $c(\mathbf{k})$ is required. We argue, however, that the DIM can even compete with analytical approximations like KC and EL in contexts where the latter are particularly well suited and in routine use, e.g.\ as part of oceanographic models where the KC approximation 
{is currently in use. }
Making as fair a comparison of these fundamentally different methods as we have been able to, we show that the 
iterative DIM algorithm
predicts $c(\mathbf{k})$ for a typical wind--driven shear profile with the same accuracy as the KC (better than $5\%$) when the $z$ axis is discretized with only $7$ points, performing just a single iteration, and using a naive and inaccurate initial value of $c$. The cost involved is of comparable magnitude to that of the analytical approximations whose integrals are evaluated with the same method as those required for the iterative DIM algorithm 
. This holds true even when including estimation of error during DIM implementation (KC is in practice used ``blindly'' with no accuracy check). Based on these cost considerations and the mentioned advantages of error control and additional cost--free flow field information it is our opinion that the DIM can compete with analytical approximations even {in such applications.}

We have applied the DIM to several examples, some of which have not been considered before to our knowledge. We make use of the DIM's ability to easily handle turning velocity profiles to compare the KC and EL approximations in this case, something which was not done in \cite{ellingsen17} due to lack of a suitable computation method. 
We secondly calculate the velocity and pressure fields beneath a wave created by a short, localized pressure pulse upon a background flow representing a near--surface shear layer, and compare them to the case without shear. Although the surface deformation is only moderately different in the two cases, the sub-surface flow field (when background flow is subtracted) is strikingly different.

We have demonstrated that the DIM can be used for efficient evaluation of the wave-action conservation equation in the presence of shear currents and slowly varying bathymetry (shear-WAC) derived by Quinn et al (2017). These authors themselves commented that this was not practical due to high cost; we {argue }
otherwise. The shear-WAC is re-written in a suitable form, and applied for demonstration to waves above a depth changing linearly between two constant levels, with an exponentially decaying surface shear current. Thus the DIM {seems to be }
a viable way in which the shear-WAC can be applied in oceanographic wave models.

\acknowledgments
YL is funded by the Faculty of Engineering, Norwegian University of Science and Technology. S{\AA}E acknowledges funding from the Norwegian Research Council (FRINATEK), project number 249740. 
We have benefited greatly from discussions with Prof Anne Kvern\o, 
  and we thank our `user panel' Peter Maxwell and Benjamin K.\ Smeltzer, for extensive testing and suggesting improvements.
No new data was generated for the research reported herein, and all equations necessary to reproduce the results are included. 

\appendix

\section{Numerical performance} \label{sec:numDem}

The DIM, concisely formulated in the coupled equations \eqref{eq:RHE} and \eqref{eq:dsp_R}, is, from a numerical point of view, a scalar system with $\tilde{c}$ as unknown. For some value of $\tilde{c}$, Eq.~\eqref{eq:RHE} implicitly defines the function $\bar{w}(z; \tilde{c})$. In an iterative scheme at iteration $n+1$, $\tilde{c}^{n+1}$ will depend on $\tilde{c}^n$ and $\bar{w}^n$, but $\bar{w}^{n+1}$ depends only on the most recent value of $\tilde{c}$, not on $\bar{w}^n$. 

The convergence of our iterative implementation of the DIM as a whole thus shares the well known criteria for Newton's method, in particular that $D_R(z)$ has continuous derivative with respect to $\tilde{c}$ at the root. It is obvious from Eqs.~\eqref{eq:DR} and \eqref{eq:ic} that this is so when there are no critical layers, i.e., when $\mathbf{k}\cdot\Delta\mathbf{U}-k\tilde{c}\neq 0$ throughout the range of $z$. Also critical layers in the interior of the range pose no problems, since it is well known that $\bar{w}(z)$ is continuous across critical layers. A critical layer cannot occur at the surface. The remaining point of interest is the case where a critical layer occurs at $z_s=-h$. In this case $\frac{\partial I_c}{\partial \tilde{c}}$ appears to have a double pole at the lower endpoint of the integral. However, the numerator of Eq.~\eqref{dIcdc} has a factor $\sinh(k(z+h))$, and the boundary condition $\bar{w}(-h)=0$ ensures that $\partial D_R/\partial \tilde{c}$ exists and is continuous also in this case. Convergence is thus assured providing the initial guess for $\tilde{c}$ is sufficiently close to the root. We have yet to come across a case where $c_0(\mathbf{k})$ is not an adequate choice for convergence. 

We have chosen a simple central difference approximation for $\bar{w}(z)$ from Eq.~\eqref{eq:RHE}, which is known to converge at least as $N^{-2}$. The same rate of convergence is true of Newton's method (c.f., e.g. Chap.3 \& 8 in \citet{isaacson12}), so an overall convergence rate of at least $N^{-2}$ is expected, and indeed found in the case considered below. 

As with many numerical integration schemes, convergence issues can arise if the grid is too course, i.e., $N$ is too small. In this case the numerical evaluation of $\bar{w}$, and hence $D_R$ and $\partial D_R/\partial c$, will have error. Cases with high values of $U''(z)$ will require a finer grid, although for typical oceanographic profiles such as in Section \ref{sec:analcomp}, convergence is fast already at $N=7$. A general criterion for the minimum value of $N$ to ensure convergence remains an open question.
 
To show how the iterative algorithm numerically converges, 
we make use of a class of special class of shear currents that satisfy $U''(z)/U(z)=$constant, analysed by \cite{peregrine76} (Section IV.B.2). Assuming $c=0$ the corresponding streamwise wave number can be found exactly. We will do the opposite: given a profile $\mathbf{U}(z)=U(z)\mathbf{e}_x$ and streamwise wave number $k_x$ for which $c=0$ is the exact solution, we use the DIM to estimate $k$ with increasing accuracy by increasing the discretization $N$ and the number of iterations, $M$.

We let $ \mathbf{U} (z) = U(z)\mathbf{e}_x$ where
\begin{equation} \label{peregrine}
U(z)=U_0\cosh \kappa z + U_0' \kappa^{-1}\sinh \kappa z.
\end{equation} 
Assuming $c=0$ (stationay waves in chosen frame of reference),
the Rayleigh equation \eqref{rayleigh} has the exact solution $w(z)=w(0)\sinh K(z+h)/\sinh K h$ with $K=\sqrt{k_x^2+\kappa^2}$. Given parameters $\kappa, U_0$ and $U_0'$, the streamwise wave number component $k_x$ solves the implicit dispersion relation $  Kh\coth Kh = gh/U_0^2 + U_0' h/U_0$. We choose $U(-h)=0$, which fixes $\kappa$ implicitly. Calculated numerical values for 
{$c $ }
will converge towards zero. We perform calculations for various propagation directions $\theta\in \langle \frac\pi 2, \frac{3\pi}2\rangle$, i.e.\ different values of $k_y$ and hence $k$. 

\begin{figure}
	\graphicspath{{figures/}}
	\centering{\includegraphics[width=\textwidth]{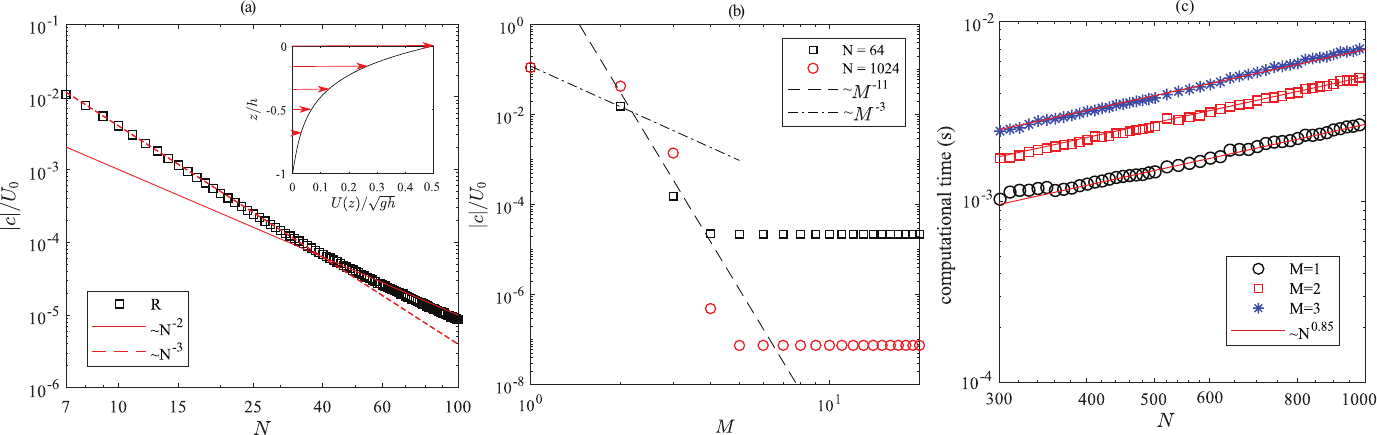}}	
	\caption{
		{
			Test cases for which $c=0$ is the exact solution: Velocity profile \eqref{peregrine} is assumed with parameters $U_0/\sqrt{gh}=0.5, hU_0'/U_0=4$, $\kappa$ given by $U(-h)=0$, whereupon, for a chosen propagation direction $\theta\in \langle\frac\pi 2,\frac{3\pi}2\rangle $, the appropriate $k$ is calculated (see main text). 
			(a) Squares: $|c|/U_0$ for increasing $N$ for $\theta=\frac{5\pi}4$. Lines are proportional to $N^{-2}$ (solid) and $N^{-3}$ (dashed). 
			(b) Squares and circles: $|c|/U_0$ for increasing $M$ for $\theta=\frac{5\pi}4$ when $ N = 64 $ and $ N=1024 $, respectively. Lines are proportional to $M^{-11}$ (dashed) and $M^{-3}$ (dash dot).
			(c) Calculational time on desktop computer for $M$ iterations, $50$ equidistant values of $k$ between $0.51\pi$ and $1.49\pi$. See main text for further details.
		}}
		\label{fig:pg}
\end{figure}

Convergence is tested for a single iteration and increasing grid refinement in Fig.~\ref{fig:pg}a {and for increasing iterations in Fig.~\ref{fig:pg}b}. We consider propagation in direction $\theta=5\pi/4$ as a representative example 
and we used a naive initial guess $c_0$ for Fig.~\ref{fig:pg}b.
The figure shows that with our implementation the convergence
with respect to $N$ is better than $\sim N^{-2}$, and approximately $\sim N^{-3}$ for the level of accuracy required in many practical applications. Even for $N=1024$ accuracy becomes limited by $N$, not $M$ already after $4$-$5$ iterations.

In Fig.~\ref{fig:pg}c we show calculation time on a standard desktop computer 
(8 processors: Intel i7-4770 3.4 MHz, 32 GB RAM) 
for $M=1, 2$ and $3$ iterations and increasing discretisation $N$. In order to test a range of different wavelengths, each calculation runs through 50 values of $k_y$ (the value of $k_x$ will be the same for all) by choosing $50$ equidistant values of $\theta$ in the range $\langle \frac\pi 2,\frac{3\pi}2\rangle$ of counterstreamwise directions, where 
stationary wave solutions are possible. In all cases we find that calculation time scales approximately as $N^{0.85}$. 

\subsection{Wind-induced profiles}

\begin{figure*}
	\graphicspath{{figures/}}
	\centering{\includegraphics[width=0.8\textwidth]{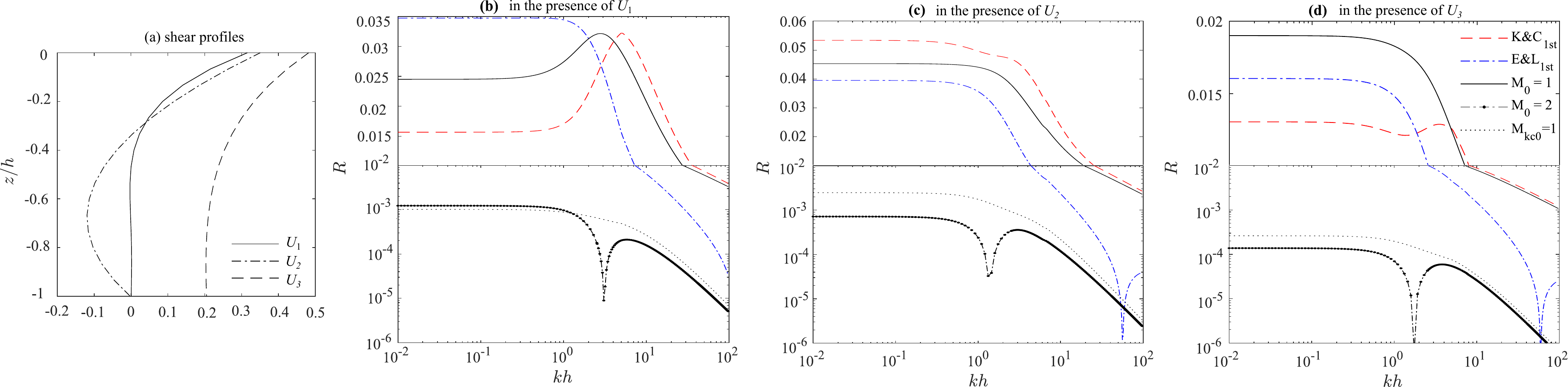}}	
	\caption{
		(a) Wind--driven velocity profiles taken from \cite{swan00}. (b,c,d) Relative errors of calculated value of $\tilde{c}(k)$ for waves on the profiles in panel a, respectively. See main text for details.	
	}
	\label{fig:wind}
\end{figure*}

In this section 				
our 
demonstration is for three typical examples of wind--induced surface flows, taken from \cite{swan00}, shown in Fig.~\ref{fig:wind}a. For five different calculation procedures we study the relative error made in the calculation, $R=|\tilde{c}_\approx-\tilde{c}_e|/\tilde{c}_e$, where $\tilde{c}_e$ is the fully converged, ``exact'' value. 
The five procedures are labelled as follows. K\&C$_\text{1st}$: Approximation \eqref{KC}. E\&L$_\text{1st}$: Approximation \eqref{EL}. $M_0=1, M_2=2$:, calculations using DIM with $1$ and $2$ iterations, respectively, using $c_0$ as the initial guess. $M_{\text{KC}0}=1$: one iteration of DIM, using approximation \eqref{KC} as initial guess. (For a careful analysis of the performance of Eqs.~\eqref{KC} and \eqref{EL} and their extensions, see \cite{ellingsen17}.)  In all methods the DIM as well as numerical integration is performed on appropriate grids of $z$-values with  $N=7$. 

Figure \ref{fig:wind} shows how a single iteration of DIM with the no--shear velocity $\tilde{c}=c_0$ as initial guess, gives results that are as good as the analytical approximations \eqref{KC} and \eqref{EL}. In the discussions in section \ref{sec:analcomp} we show that calculation times are also typically similar, although depending on context and exact implementation. In all cases a second iteration of the DIM brings the calculated result to within 
$0.1\%$ (much better in most of the cases)
of the true value, more than adequate for 
many practical purposes. The same high accuracy or better is obtained with a single iteration of the DIM if the analytical approximation \eqref{KC} is used as initial guess. 
A second iteration of DIM is much faster than including the second order accurate analytical expressions using \citet{kirby89} or \citep{ellingsen17}, which give similarly high accuracy.

\listofchanges

%
\end{document}